\documentclass[a4paper,10pt]{article} 
\usepackage{cite}
\usepackage{setspace}
\usepackage{ae,aecompl,amsbsy,amssymb,eurosym}
\usepackage[english]{babel}
\usepackage{graphicx}
\usepackage{epsfig}
\usepackage{epstopdf}
\usepackage{amsmath}
\usepackage{amssymb}
\usepackage{subfigure}
\usepackage{ifthen}
\usepackage{alltt}
\usepackage{fancyhdr}
\usepackage{verbatim}
\usepackage{moreverb}
\usepackage{url}

\newcommand{\vt}[1]{\ensuremath{\boldsymbol{#1}}} 
\newcommand{\lt}[1]{\ensuremath{\mathrm{#1}}} 
\newcommand{\ec}{\ensuremath{\vt{J}}} 
\newcommand{\mc}{\ensuremath{\vt{M}}} 

\usepackage[pdftex,bookmarks,colorlinks=false]{hyperref}

\fancypagestyle{plain}{
  \fancyhead{} 
}

\begin{document}

\author{A.~O.~Karilainen, P.~Ikonen, C.~R.~Simovski, and S.~A.~Tretyakov}
\title{Choosing Dielectric or Magnetic Material to Optimize the Bandwidth of Miniaturized Resonant Antennas}
\maketitle

\begin{abstract}


We address the question of the optimal choice of loading material for antenna miniaturization. A new approach to identify the optimal loading material, dielectric or magnetic, is presented for resonant antennas. Instead of equivalent resonance circuits or transmission-line models, we use the analysis of radiation to identify the fields contributing mostly to the stored energy. This helps to determine the beneficial material type. The formulated principle is qualitatively illustrated using three antenna types. Guidelines for different antenna types are presented.

\end{abstract}


\section{Introduction}

Demand towards smaller and smaller portable communication devices has continuously challenged antenna engineers to come up with solutions to make the antennas smaller. However, the task has not been trivial since especially bandwidths of antennas are not always allowed to decrease. Resonant antennas, the type used e.g.\ in most if not all hand-held devices, can be made to operate at lower frequencies by making their electric lengths larger. This can be easily achieved by inserting some material to the structure so that the resonating wavelength decreases. Of course, when the fields in the resonators are changed, the operation of the antenna is not only changed in view of the resonance frequency but also the bandwidth, matching, efficiency, etc.\ are affected.  Miniaturization of patch antennas using materials with various material parameters $\epsilon$ and $\mu$ studied in \cite{Hansen2000} aroused discussion of the benefits of different materials with magnetic response. Various loading scenarios have been considered, for example, dielectric coated dipole antennas were experimentally and theoretically studied in \cite{Lamensdorf1967}, dipole antennas covered with metamaterials were studied in \cite{Ziolkowski2003}, and double negative materials for electrically small antennas in \cite{Tretyakov2005}. In this article, we discuss how to determine the proper loading material for a small resonant antenna. Especially, we focus on self-resonant antennas which do not require matching networks. The analysis here assumes lossless non-dispersive materials, because the goal is to determine the optimal loading type (dielectric or magnetic). Effects of material dispersion (reduced bandwidth) and losses (reduced efficiency) are well known.

Metamaterials with tailored magnetic response have been under study as well as magneto-dielectric materials with natural magnetic response. Magneto-dielectric meta-substrates (metamaterials) were used to miniaturize a patch antenna in \cite{Mosallaei2004}, \cite{Mosallaei2004b}. A patch antenna with metamaterial substrate was further studied with the dispersion included, and challenged against reference antennas with dielectric substrates in \cite{Ikonen2006b}. It was theoretically and experimentally seen that the negative effect of strong dispersion is stronger than the beneficial effect of the artificial magnetic substrate. Magneto-dielectric substrates with natural magnetic inclusions were however seen to provide wider bandwidths than reference dielectric substrates \cite{Ikonen2006c}.  The reduced stored energy while preserving the radiation fields in patch antennas with ideal dielectric and magnetic substrates was clearly illustrated in \cite{Ikonen2008}. The explanation for this effect was seen in that the fringing electric field at patch slots is the main radiation mechanism, while the magnetic field in the antenna vicinity contributes mostly to the stored energy. Here, we extend the approach established in \cite{Ikonen2008} and propose a general rule for determination of the beneficial substrate type from the analysis of the radiation mechanism.

Resonance circuits are sometimes used to model the behavior of resonant antennas near the resonance frequency. An equivalent ladder circuit model for small omni-directional antennas was used in \cite{Chu1948}. For finite sized dipole antennas equivalent circuits are discussed e.g.\ in \cite{Tang1993} and \cite{Hamid1997}. Patch antenna is analyzed in \cite{Ikonen2006c}. Below we show that equivalent circuits are not always to be trusted when considering antenna miniaturization.

In the following sections, we first discuss the equivalent circuit models: how the material loading affects the antenna bandwidth and what is the role of the feeding type. Also, the use of the transmission-line theory is discussed with material loading in mind. Next, we introduce the original field approach and formulate rules. Finally, we present examples with a patch antenna, a dipole antenna, and a planar inverted-L antenna (PILA).


\section{Antenna miniaturization modeled using resonance equivalent circuits and transmission lines}

When discussing miniaturization of small resonant antennas, the quality factor $Q$ plays an important role, and it is written as:
\begin{equation}
	Q = \frac{\omega W}{P_\lt{rad}},
	\label{eq:Q}
\end{equation}
where $\omega$ is the angular frequency, $W$ is the average stored energy and $P_\lt{rad}$ is the radiated power. Since the bandwidth $B$ is inversely proportional to the quality factor:
\begin{equation}
	B \sim \frac{1}{Q},
	\label{eq:B}
\end{equation}
a small $Q$ is usually desired. The decrease in $Q$ can be achieved in different ways. Ideally, the stored energy $W$ should be minimized, while the radiated power $P_\lt{rad}$ should be maximized.


\subsection{Equivalent circuits for series and parallel type resonances}

The antenna quality factor can be described also in terms of the circuit theory: The slope of the reactive part $X_\lt{a}$ of the antenna impedance $Z_\lt{a} = R_\lt{a} + jX_\lt{a}$ should be minimized, while the desired losses due to the radiation resistance $R_\lt{a}$ are not decreased (assuming no losses due to resistive dissipation to heat), according to (e.g.,~\cite{Ikonen2006b})
\begin{equation}
	Q = \frac{\omega}{2R_\lt{a}}\frac{\partial X_\lt{a}}{\partial \omega}
	\label{eq:Q_from_Z}
\end{equation}
in case of a series-type resonance of $Z_\lt{a}$. One has to remember that the change in radiating fields and hence in $P_\lt{rad}$ affects the radiation resistance $R_\lt{a}$, which has to be considered when the antenna is connected to the feeding waveguide.

\begin{figure}[!t]
  \centering
  \subfigure[]
  {
    \includegraphics{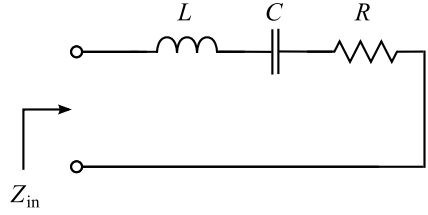}
    \label{fig:series_rlc_circuit}
  }
  \subfigure[]
  {
    \includegraphics{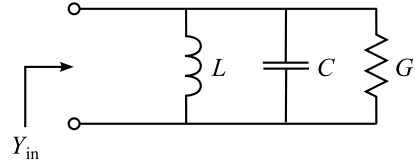}
    \label{fig:parallel_rlc_circuit}
  }
  \caption{a)~Series and b)~parallel $RLC$ resonant circuit.}
  \label{fig:rlc_circuits}
\end{figure}

The main reason that the circuit theory analogy has been widely used (e.g.\ in \cite{Ikonen2006c}) is its simplicity. The input impedance $Z_\lt{in}$ or admittance $Y_\lt{in}$ and the quality factor of a resonator can be neatly written using a parallel or series type $RLC$ circuit in the vicinity of the resonances, see Fig.~\ref{fig:rlc_circuits}. Let us inspect both of the resonator types separately (see e.g.~\cite{microwave_engineering}).

At series resonance, the quality factor for the series resonance $RLC$ circuit of Fig.~\ref{fig:series_rlc_circuit} is
\begin{equation}
  Q^\lt{ser} = \frac{\omega_{0}L}{R} = \frac{\omega_{0}}{\Delta\omega^\lt{ser}}  =  \frac{1}{\omega_{0}RC},
  \label{eq:Q_series}
\end{equation}
where $\omega_0$ is the resonance angular frequency and $\Delta\omega^{\mathrm{ser}}$ is the half-power bandwidth. Similar equation can be derived from the parallel $RLC$ circuit shown in Fig.~\ref{fig:parallel_rlc_circuit}:
\begin{equation}
  Q^\lt{par} = \omega_{0}RC = \frac{\omega_{0}}{\Delta\omega^\lt{par}} = \frac{R}{\omega_{0}L}.
  \label{eq:Q_parallel}
\end{equation}
The resonance angular frequency for both circuits is
\begin{equation}
  \omega_{0} = \frac{1}{\sqrt{LC}}.
  \label{eq:undamped_resonance_frequency}
\end{equation}

Now, let us assume that a resonant antenna behaves like one of these resonators in the vicinity of its resonance frequency. When we load or fill the antenna with a magnetic material, it is analogous to filling the coils in Fig.~\ref{fig:series_rlc_circuit} and Fig.~\ref{fig:parallel_rlc_circuit} with e.g. a ferrite core. Let us assume further, that the increase in inductance $L$ is compensated accordingly by decreasing the capacitance $C$ in (\ref{eq:undamped_resonance_frequency}), so that we maintain the same resonance angular frequency $\omega_0$. The results can be seen clearly for the series resonator from (\ref{eq:Q_series}): The quality factor $Q^\lt{ser}$ increases and the bandwidth $\Delta\omega^\lt{ser}$ decreases. For the parallel resonator the effect is opposite, as seen from (\ref{eq:Q_parallel}). As $C$ is decreased, the quality factor decreases and the bandwidth increases. In other words, if an antenna operating at the parallel resonance can be loaded (completely or partially filled) with a magnetic material, the size decrease does not harm the bandwidth, opposite to the dielectric loading.

The role of the feeding has to be at the same time taken into account. For example, a small electric dipole is fed so that the complex amplitude of the input current $I$ is fixed. Then, by loading the dipole with a dielectric material so that the electric field is reduced, we decrease the average stored electric energy through decreased voltage $V$~\cite{microwave_engineering}:
\begin{equation}
	W_\lt{e} = \frac{1}{4}|V|^2 C,
	\label{eq:RLC_We}
\end{equation}
while keeping the magnetic energy
\begin{equation}
	W_\lt{m} = \frac{1}{4}|I|^2 L
	\label{eq:RLC_Wm}
\end{equation}
constant since $I$ is fixed. An analogy follows for a small magnetic dipole, which can be thought to be a small electric current loop. An $RLC$ circuit equivalent is the parallel connection of Fig.~\ref{fig:parallel_rlc_circuit}, where the voltage is high since $C$ and $L$ form an open circuit this time. If we maintain the voltage driving the loop, the magnetic field and $W_\lt{m}$ can be decreased with magnetic material while $W_\lt{e}$ is maintained.

The main drawback of this $RLC$ description of antennas is as follows: An antenna impedance dispersion changes with simple impedance transformer and moreover with a matching network while the antenna remains the same. Also, this simple theory assumes that material loading does not change the current mode nor the radiation resistance $R_\lt{a}$ of the antenna at the resonance frequency. Our approach to the description of antenna loading will be free of this vagueness.

\subsection{Transmission-line resonators}
\label{sec:tl_resonators}

\begin{figure}[!t]
  \centering
  \includegraphics{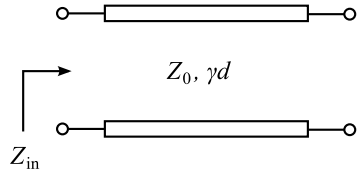}
  \caption{A transmission-line resonator with open circuit termination. The transmission line has the characteristic impedance $Z_0$ and the electric length $\beta d$. The propagation constant is $\gamma=\alpha+j\beta$.}
  \label{fig:tl_circuit_open}
\end{figure}

Transmission-line (TL) resonators also behave like series and parallel resonators near the resonance frequencies~\cite{microwave_engineering}. In fact, we can find similar equivalent values of $R$, $L$ and $C$ for resonating TLs. For example, we can calculate the input impedance $Z_\lt{in}$ for a TL of the length $d=\lambda/2$ with an open-circuit termination, as shown in Fig.~\ref{fig:tl_circuit_open}. The TL is assumed to be lossy with non-zero attenuation coefficient $\alpha$ of the complex propagation constant $\gamma = \alpha + j\beta$ so that the quality factor is finite. It turns out that the impedance behaves like the parallel equivalent circuit in Fig.~\ref{fig:parallel_rlc_circuit}, and the quality factor can be solved as:
\begin{equation}
	Q_{\lambda/2}^\lt{par} = \omega_0 RC = \frac{\pi}{2 \alpha d} = \frac{\beta}{2\alpha},
	\label{fig:Q_half-lambda}
\end{equation}
where $R=Z_0/(\alpha d)$ and $C=\pi/(2\omega_0Z_0)$ are the equivalent resonance circuit element values and $Z_0$ is the characteristic impedance of the TL. The resonance frequency and $L=1/(\omega_0^2C)$ is of course determined by (\ref{eq:undamped_resonance_frequency}).
If the length of the TL with the same open circuit termination is only $d=\lambda/4$, we can similarly derive $Z_\lt{in}$. Now we see that it behaves like a series resonance circuit of Fig.~\ref{fig:series_rlc_circuit} and the quality factor is
\begin{equation}
	Q_{\lambda/4}^\lt{ser} = \frac{\omega_0L}{R} = \frac{\pi}{4 \alpha d} = \frac{\beta}{2\alpha},
	\label{fig:Q_quarter-lambda}
\end{equation}
where $R=Z_0\alpha d$, $L=Z_0\pi/(4\omega_0)$ and $C=1/(\omega_0^2L)$.

If we consider the TL as a TEM waveguide, it is now hard to see how the material filling parameters would affect the quality factors, since $\beta=k_0\sqrt{\mu_\lt{r}\epsilon_\lt{r}}$ is proportional to both material parameters $\mu=\mu_\lt{r}\mu_0$ and $\epsilon=\epsilon_\lt{r}\epsilon_0$ of the medium, as seen in (\ref{fig:Q_half-lambda}) and (\ref{fig:Q_quarter-lambda}). In other words, the general transmission-line theory does not describe the antenna geometry through $Z_0$, not to mention the radiation in enough detail, so that definite conclusions could be made. More detailed antenna models, which describe the fields and especially the radiation, must be used.


\section{Antenna miniaturization modeled using equivalent radiating currents}

From the resonant-circuit model of small resonant antennas as well from the transmission-line model (the previous section)
it appears that the choice of the material loading type (dielectric or magnetic) is fully determined by the resonance type (series or parallel). 
However, there are antennas that apparently do not obey this rule. One of such antennas is the planar inverted-L antenna (described below).
This calls for developing a more advanced approach to this problem. 
The challenge is to describe the matching-independent effect of decreasing the stored energy (or enhancing the radiation) in real antennas, which can be achieved by the material loading. Because real antennas have three-dimensional structures, various shapes and field distributions, the one-dimensional equivalent circuit or transmission-line representation are not accurate enough. We will see that quite often for resonator antennas one field (electric or magnetic) provides the main contribution to the radiation, whereas the other field contributes mainly to the reactive energy storage. The strategy for miniaturization of an antenna is then to fill or cover the antenna so that the second field is suppressed, whereas the  radiating field is not diminished.

Let us first formulate the problem. The antenna in volume $V$ can be enveloped by a surface $S$, and the equivalent surface electric and magnetic currents $\ec_\lt{s}$ and $\mc_\lt{s}$ that produce the radiated field can be calculated from Huygens' principle (the equivalence principle, see e.g.~\cite{harrington}) using the tangential components of electric and magnetic fields $\vt{E}$ and $\vt{H}$, respectively, on $S$:
{\setlength\arraycolsep{2pt}
\begin{eqnarray}
  \ec_\lt{s} & = & \vt{n} \times \vt{H} \big|_S,
    \label{eq:electric_eq_current}
    \\
  \mc_\lt{s} & = & -\vt{n} \times \vt{E} \big|_S.
  \label{eq:magnetic_eq_current}
\end{eqnarray}
Here $\vt{n}$ is the unit normal vector pointing outside from $V$. Now the total  fields outside $V$ can be calculated from $\ec_\lt{s}$ and $\mc_\lt{s}$ on $S$ using the vector potentials \cite{balanis}
{\setlength\arraycolsep{2pt}
\begin{eqnarray}
  \vt{A}(\vt{r}) & = & \frac{\mu}{4\pi} \int_S \ec_\lt{s} (\vt{r}') \frac{e^{-jk|\vt{r}-\vt{r}'|}}{|\vt{r}-\vt{r}'|} \lt{d} S',
    \label{eq:field_from_electric_eq_current}
    \\
  \vt{F}(\vt{r}) & = & \frac{\epsilon}{4\pi} \int_S \mc_\lt{s} (\vt{r}') \frac{e^{-jk|\vt{r}-\vt{r}'|}}{|\vt{r}-\vt{r}'|} \lt{d} S',
  \label{eq:field_from_magnetic_eq_current}
\end{eqnarray}
where $k$ is the wavenumber and the variables marked with the apostrophe correspond to the source point. The fields in whole space can be then written as~\cite{balanis}
{\setlength\arraycolsep{2pt}
\begin{eqnarray}
  \vt{E} & = & -j \omega \vt{A} - j\frac{1}{\omega\mu\epsilon} \nabla ( \nabla \cdot \vt{A} ) - \frac{1}{\epsilon} \nabla \times \vt{F},
    \label{eq:EfromAF}
    \\
  \vt{H} & = & -j \omega \vt{F} - j\frac{1}{\omega\mu\epsilon} \nabla ( \nabla \cdot \vt{F} ) + \frac{1}{\mu} \nabla \times \vt{A}.
  \label{eq:HfromAF}
\end{eqnarray}

If an antenna radiates mostly from either electric field ($\mc_\lt{s}$) or magnetic field ($\ec_\lt{s}$) according to equations (\ref{eq:electric_eq_current}) and (\ref{eq:magnetic_eq_current}), the non-radiating field can be modified using magnetic or dielectric material loading, respectively, without significantly affecting the radiation. The modification should be of course chosen so that the field is diminished with e.g.\ magnetic or dielectric material. In other words, we find that the beneficial loading material for resonant antennas is determined by the type of the main radiating (electric of magnetic) current on $S$.


\section{Examples and guidelines}

\subsection{Dipole antenna}

 Let us consider a monopole over an infinite ground plane or a dipole antenna, as seen in Fig.~\ref{fig:dipole_covered}. This structure radiates mostly from the surface current $\ec$, or from the equivalent current created by the magnetic field: $\ec_\lt{s} = \vt{n} \times \vt{H}$. The electric field $\vt{E}$ is normal at PEC cylinder and therefore mostly normal to $S$ located close to the antenna (see Fig.~\ref{fig:dipole_covered}), and $\mc_\lt{s} \ll \ec_\lt{s}$. Equivalent electric surface current $\ec_\lt{s}$ caused by the magnetic field is preserved, when the volume $V$ bounded by $S$ is filled with a dielectric material. Therefore, miniaturization with dielectric loading is possible without disturbing the radiating electric current. The magnetic loading leads to the miniaturization simultaneously with the decrease of bandwidth. In practice, the bandwidth also somewhat decreases with a dielectric coating \cite{Lamensdorf1967}, but this effect is much stronger with magnetic coating. The antenna impedance resonance of a dipole antenna in its fundamental mode is of the series type, and the antenna behaves as expected in terms of miniaturization according to the equivalent circuit model.

\begin{figure}[!t]
  \centering
  \includegraphics[width=85mm]{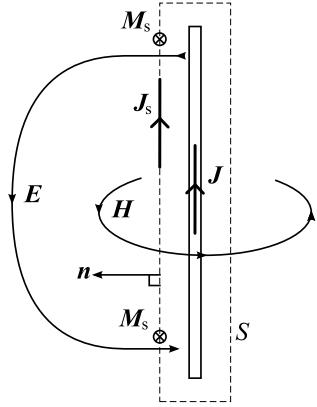}
  \caption{Dipole antenna with the surface $S$ brought away from the surface of the antenna.}
  \label{fig:dipole_covered}
\end{figure}

\subsection{Patch antenna}

For an opposite example, assume a simple cavity model for the patch antenna with a PEC ground plane and patch, as seen in Fig.~\ref{fig:patch_cavity}. By neglecting the fringing fields using the PMC boundary condition at the aperture surrounding the cavity, we have vertical electric field across the sidewalls of the cavity, and using (\ref{eq:magnetic_eq_current}) we have magnetic equivalent surface current $\mc_\lt{s} = -\vt{n} \times \vt{E}$ at the walls~\cite{balanis}. The surface current density $\vt{J}$ in the top patch is relatively weak when the patch height is small. Also, $\ec$ is parallel to the ground plane, so the image current is out of phase and hence the electric surface current contributes weakly to radiation.

Since the surface current does not produce much radiation, this structure radiates mostly from the magnetic equivalent current $\mc_\lt{s}$ at the aperture. Now, magnetic filling can be used for miniaturization without affecting the radiating electric field. As it happens, in case of a $\lambda/2$ patch antenna, the electric fields at the patch ends actually do not suffer at all from magnetic filling when compared to dielectric filling, as shown in \cite{Ikonen2008} in theory and with simulation. The current and hence the magnetic field under the patch is on the other hand reduced. Unlike the previous case, the optimal material loading not only miniaturizes (reduces the wavelength of the resonance mode) but also increases the relative bandwidth. Again, as the antenna impedance (neglecting the reactance of the probe) behaves as a parallel circuit, the equivalent circuit model gives the same conclusion.

\begin{figure}[!t]
  \centering
  \includegraphics[width=85mm]{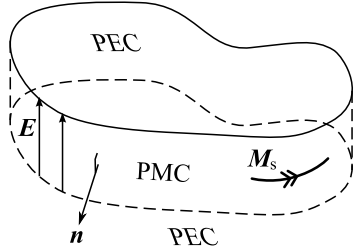}
  \caption{Cavity model of a patch antenna.}
  \label{fig:patch_cavity}
\end{figure}

\subsection{Planar inverted-L antenna}

A planar inverted-L antenna (PILA) is a quarter-wavelength patch antenna with a series-type resonance. This behavior can be understood from a simple open-ended transmission line (see Section~\ref{sec:tl_resonators}). The real part of the input impedance of a lossless transmission line is nearly zero, and the imaginary part crosses zero at the resonance, just like in the case of a $\lambda/2$ dipole antenna. However, it will be shown below that when the permittivity of a dielectric substrate is comparable with the permeability of a magnetic substrate, the magnetic response of the substrate increases the relative bandwidth.

We will study a patch antenna at the frequency where its length is $\lambda_{\rm eff}/4$. Here $\lambda_{\rm eff} = 2\pi/\beta_\lt{eff}$ is the effective wavelength in the patch and $\beta_{\rm eff}$ is the effective wavenumber. The antenna is illustrated in Fig.~\ref{fig:sketch}. We will consider the case when the distance from the edge to the feed point $a \ll d$, and let the inductance and resistance of a presumed feeding pin be negligible. We also neglect the conductivity losses and losses in the substrate (modeled by material parameters $\epsilon_\lt{r},\ \mu_\lt{r}$).

The antenna admittance $Y_\lt{a}$ at the feed point is the sum of the radiation conductance $G$ of the left edge and the input admittance of the microstrip line with the length $d$ loaded by the same radiation conductance $G$:
\begin{equation}
	Y_\lt{a} = \frac{1}{Z_\lt{a}} = G \left( 1 + \frac{1+jY_0 \tan \beta d/G}{1 + jG \tan \beta d/Y_0} \right).
	\label{eq:y}
\end{equation}

\begin{figure}[!t]
	\centering
	\includegraphics[width=75mm]{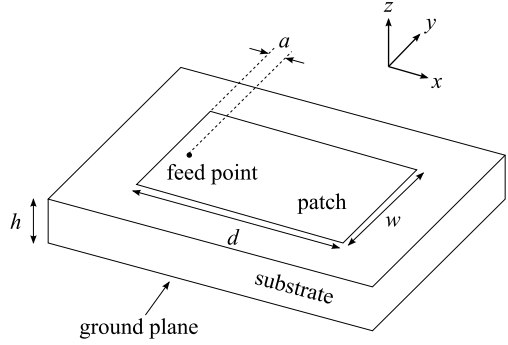}
	\caption{Illustration of a planar inverted-L patch antenna on a substrate with ground plane below.}
	\label{fig:sketch}
\end{figure}

Here $G$ is the radiation conductance of the slot, $Y_0=(w/h)\sqrt{\epsilon/\mu}$ is the characteristic admittance of the line and $\beta$ is related with the free-space wavenumber as	$\beta \approx k_0 \sqrt{ \epsilon_\lt{r}\mu_\lt{r} }$. We can write the radiation quality factor directly, when we assume lossless antenna materials, through the real and imaginary part of the antenna impedance $Z_\lt{a} = R_\lt{a} + jX_\lt{a}$ from (\ref{eq:Q_from_Z}). We set $\omega_0$ as the frequency at which $X_\lt{a}=0$ and the patch is in the $\lambda_\lt{eff}/4$ (series-type) resonance. Substituting (\ref{eq:y}) into (\ref{eq:Q_from_Z}) and applying for the derivative the resonance condition $\beta d = \pi/2$ we easily derive the radiation quality factor
\begin{equation}
	Q = \frac{w\omega_0 d}{G h} \epsilon.
	\label{eq:q1}
\end{equation}
The permeability of	the substrate cancels out and does not enter into this expression. A useful intermediate result in this derivation is as follows:
\begin{equation}
	R_\lt{a} \big|_{\omega=\omega_0} = \frac{h^2\mu}{w^2\epsilon}G.
	\label{eq:r1}
\end{equation}

In papers \cite{Pinhas1988,Derneryd1976} it was indicated (and confirmed	by the validation of the whole model) that the radiation conductance of an edge of microstrip antennas weakly depends on its length $d$. There is no practical difference between $G$ of a standard $\lambda/2$ patch antenna (experiencing the parallel resonance) and $G$ of a $\lambda/4$-patch antenna. In both cases the distribution of $E_z$ (i.e. the voltage distribution) along the $y$ axis (see in Fig.~\ref{fig:sketch}) is decoupled with the voltage distribution along $x$. In other words, $G$ depends only on the electrical width	of the slot $k_0w$.

The general formula for arbitrary $w$ was accurately derived in	\cite[formula (40)]{Jackson1991}. For $w\ll \lambda_{\rm eff}$ and $h\ll w$ this involved formula yields to the standard radiation conductance of an electrically short narrow slot in a PEC plane:
\begin{equation}
	G = A \left( \frac{w}{\lambda_0} \right)^2, \qquad A = \frac{1}{90}, \quad \lambda_0 \equiv \frac{2\pi}{k_0}.
	\label{eq:g}
\end{equation}
This approximate formula works up to $w=\lambda_0/4$ \cite{Pinhas1988}. For wider patches\footnote{The current distribution corresponding to the series resonant mode can be excited along the $X$ axis for $w>d$ when $a\ll d$.} the factor $A=1/90$ should be corrected, however it is important that it never depends on the material parameters of the substrate \cite{Derneryd1976,Jackson1991}.

Substituting (\ref{eq:g}) into (\ref{eq:q1}) we come to the result for the radiation quality factor of a $\lambda/4$-patch antenna:
\begin{equation}
	Q = \frac{2\pi d \lambda_0}{A\sqrt{\epsilon_0\mu_0}wh} \epsilon_\lt{r}
= \frac{4 \pi^2 d}{A\mu_0 wh\omega_0}\epsilon_\lt{r} .
	\label{eq:sec}
\end{equation}
Let us now compare two antennas of a same size, miniaturized using a magnetic substrate $\mu_\lt{r}=n^2,\ \epsilon_\lt{r}=1$ and a dielectric one $\epsilon_\lt{r}=n^2,\ \mu_\lt{r}=1$ (here $n>1$ can be treated as the refraction index of the substrate). In the presented model both miniaturized antennas experience the series resonance at the same frequency. The ratio of the radiation quality factors of the antennas with dielectric and magnetic substrates, i.e.\ the inverse ratio for their bandwidth follows from (\ref{eq:sec}):
\begin{equation}
	Q^\lt{rel} \equiv  \frac{Q^\lt{magn}}{Q^\lt{diel}} = \frac{B^\lt{diel}}{B^\lt{magn}} = \frac{1}{n^2} < 1,
	\label{eq:rat}
\end{equation}
where $Q^\lt{magn}$ and $B^\lt{magn}$ are the quality factor and bandwidth of the antenna with the magnetic substrate and $Q^\lt{diel}$ and $B^\lt{diel}$ are respective values for the dielectric substrate. This shows that the magnetic substrate should always provide lower $Q$ than the dielectric one, despite the fact that the antenna works in the series resonance.

This results is similar as the relative radiation quality factor $Q_\lt{r}^\lt{rel}=1/\mu_\lt{r}$ derived for a $\lambda/2$ patch in \cite{Ikonen2006b,Ikonen2006c,Ikonen2007c}. By increasing $\mu_\lt{r}$, $Q_\lt{r}^\lt{rel}=1/\mu_\lt{r}$ was below unity, i.e.\ the magnetic substrate provided lower $Q_\lt{r}$. The same thing is seen in (\ref{eq:sec}), where $Q$ is directly proportional to $\epsilon_\lt{r}$, not to $\mu_\lt{r}$.

Simulations were made to validate the presented model for the $\lambda/4$ PILA. A structure similar to Fig.~\ref{fig:sketch} was studied. The dimensions of the patch were $w=d=42$~mm, $h=5$~mm and the ground plane and the patch were modeled as PEC boundaries, zero-volume in the case of the patch. First the structure with either $\epsilon_\lt{r}$ or $\mu_\lt{r}$ substrate was simulated using Ansoft HFSS full-wave simulator \cite{HFSS11} and the quality factor was calculated using (\ref{eq:Q_from_Z}). The used equation does not rely on any specific impedance-matching condition, so in reality the low $R_\lt{a}$ needs to be matched to e.g.\ $Z_0=50~\Omega$, which in its turn affects the total $Q$ due to the stored energy in the matching network (\emph{and may change the resonance type of the input impedance}). When the $\lambda/4$ resonance frequency with $X_\lt{a} \approx 0$ was found, the realized $\omega_0$ and the chosen $\epsilon_\lt{r}$ were substituted in (\ref{eq:sec}) and the theoretical $Q$ was calculated.

The results for the aforementioned dimensions are presented in Fig.~\ref{fig:model_comparison}. The simulated and theoretical results do not give the same results in magnitude, but the difference between the $\epsilon_\lt{r}$ and $\mu_\lt{r}$-loaded antennas is clear when the material parameters change. As the permittivity of the substrate is increased, $Q$ start to increase, but as the permeability increases, $Q$ holds at the same level, as predicted by the theory.

The difference between the simulated and theoretical results are due to the approximations made in the derivation, e.g.\ the approximation  $\beta\approx k_0\sqrt{\mu_\lt{r}\epsilon_\lt{r}}$. In fact, the distributions of the electric and magnetic fields across the patch are strongly different. As a result, the ratio of the effective permittivity of the microstrip line formed by the patch and the physical substrate $\epsilon_\lt{r}$ is different from the corresponding ratio for the effective permeability and  $\mu_\lt{r}$. Therefore, the contribution of $\epsilon_\lt{r}$ into $n$ differs from that of $\mu_\lt{r}$. This factor is not taken into account in our approximate theory above.

\begin{figure}[!t]
	\centering
	\includegraphics[width=85mm]{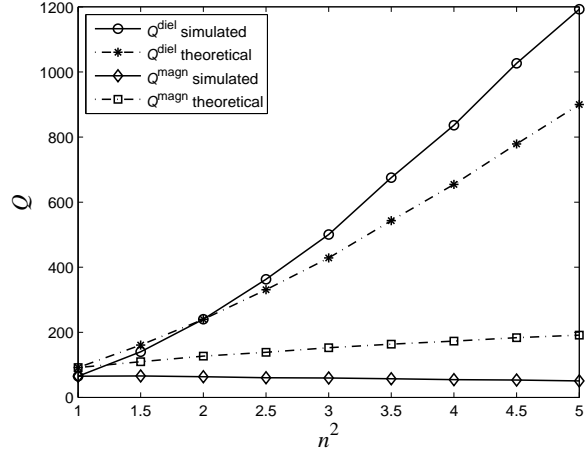}
	\caption{Simulated and theoretical quality factors for a $\lambda/4$ patch. Simulations for dielectric and magnetic substrates, using (\ref{eq:Q_from_Z}) to calculate the respective quality factors $Q^\lt{diel}$ and $Q^\lt{magn}$. Theoretical quality factors were obtained using (\ref{eq:sec}) with the realized resonance frequencies $\omega_0$ from the simulations. The reader should note, that the realized resonance frequencies $\omega_0$ are in general different with dielectric and magnetic substrates, i.e.\ $Q^\lt{diel}$ and $Q^\lt{magn}$ are not comparable with the same $n$.}
	\label{fig:model_comparison}
\end{figure}

What happens in the antenna with dielectric and magnetic loading that explains the seen behavior? Let us examine the field and current distributions in the simulated antenna with substrates of $\mu_\lt{r}=3.5$ and $\epsilon_\lt{r}=4.0$ as they produce approximately the same $\omega_0$. Fig.~\ref{fig:pila_efield} shows maximum values  of the electric field strength $|\vt{E}|$ and Fig.~\ref{fig:pila_jsurf} the surface current (with 90 degrees phase shift from $\vt{E}$) $|\vt{J}_\lt{s}|$ at the patch level. As opposed to the half-wavelength patch antenna analyzed in \cite{Ikonen2008}, now we see that $|\vt{E}|$ is higher with magnetic loading but $|\vt{J}_\lt{s}|$ is virtually the same. This difference in $|\vt{E}|$ can be understood from (\ref{eq:r1}), where the radiation resistance is proportional to $\mu$ but inversely proportional to $\epsilon$. The electric field contributes to the radiated power in (\ref{eq:Q}) through (\ref{eq:magnetic_eq_current}), and in case of the magnetic loading this overcomes the change in stored energy $W$ and we see the beneficial behavior in $Q$. This can be seen also in the antenna impedance, which is plotted in Fig.~\ref{fig:za_example}. The higher electric field is seen as the higher $R_\lt{a}$ for the magnetic filling. Although the slope of $X_\lt{a}$ and average stored energy are higher for the magnetic filling,  but not enough higher to destroy the improvement in $Q$ due to enhanced radiation.

\begin{figure}[!t]
  \centering
  \subfigure[]
  {
    \includegraphics[width=85mm]{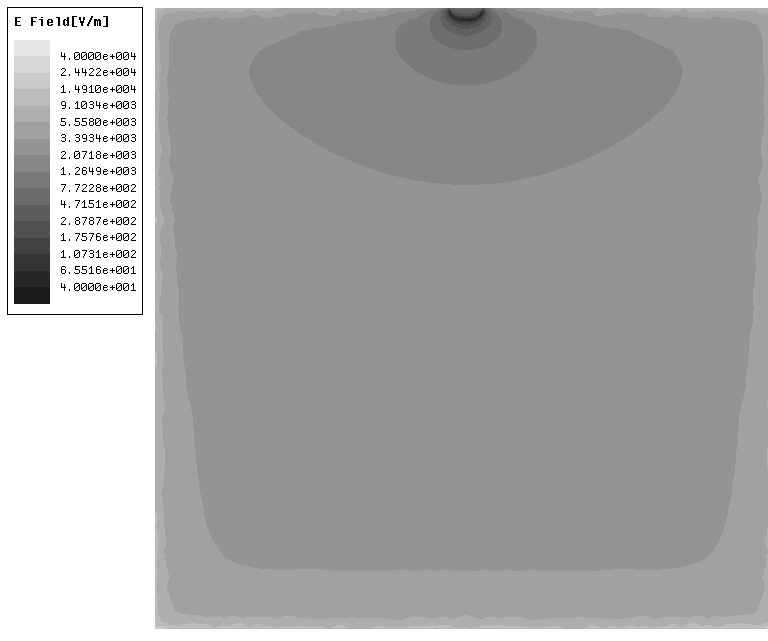}
    \label{fig:pila_diel_efield}
  }
  \subfigure[]
  {
    \includegraphics[width=85mm]{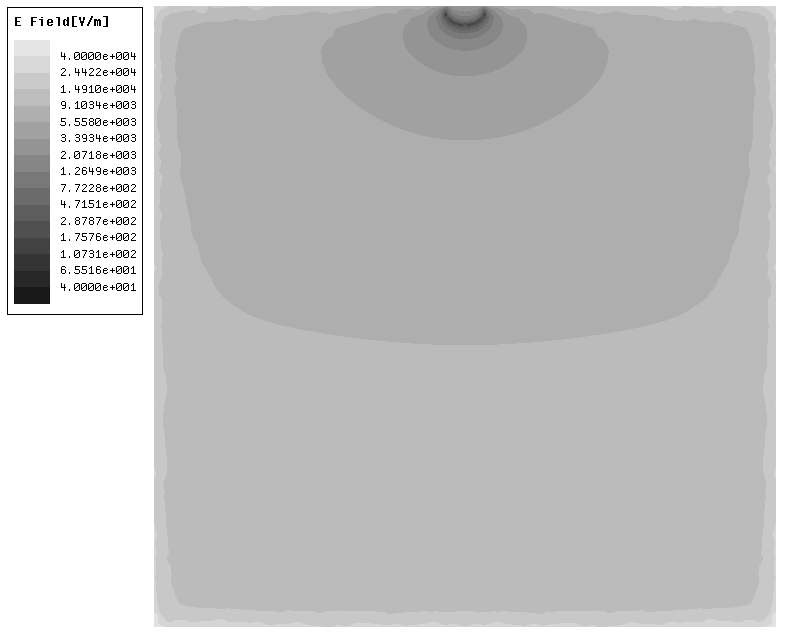}
    \label{fig:pila_magn_efield}
  }
  \caption{The electric field distributions calculated from the square $\lambda/4$ patch with a)~dielectric loading with $\epsilon_\lt{r}=4.0$ and b)~magnetic loading with $\mu_\lt{r}=3.5$, both at 550~MHz. The feed is at the center of the top edge of the patch. The electric field at the patch level is stronger at all edges with the magnetic loading.}
  \label{fig:pila_efield}
\end{figure}

\begin{figure}[!t]
  \centering
  \subfigure[]
  {
    \includegraphics[width=85mm]{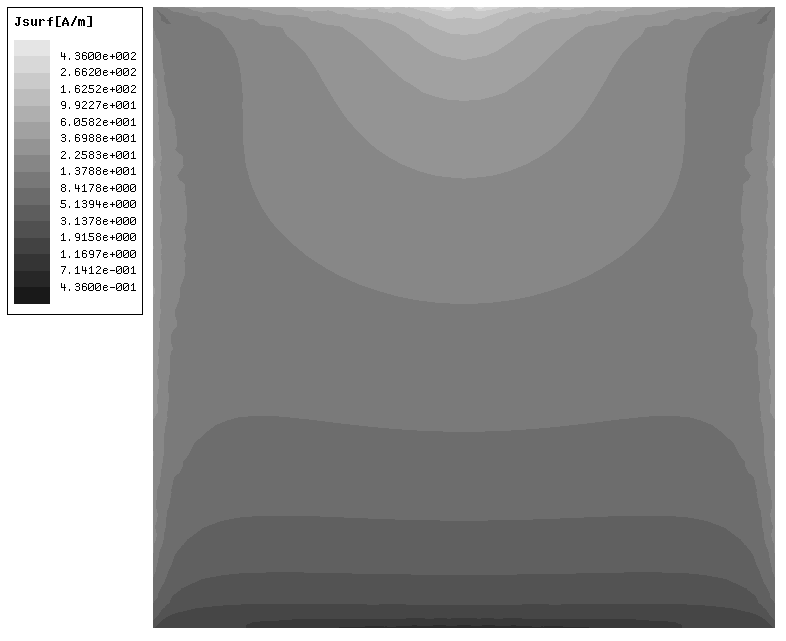}
    \label{fig:pila_diel_jsurf}
  }
  \subfigure[]
  {
    \includegraphics[width=85mm]{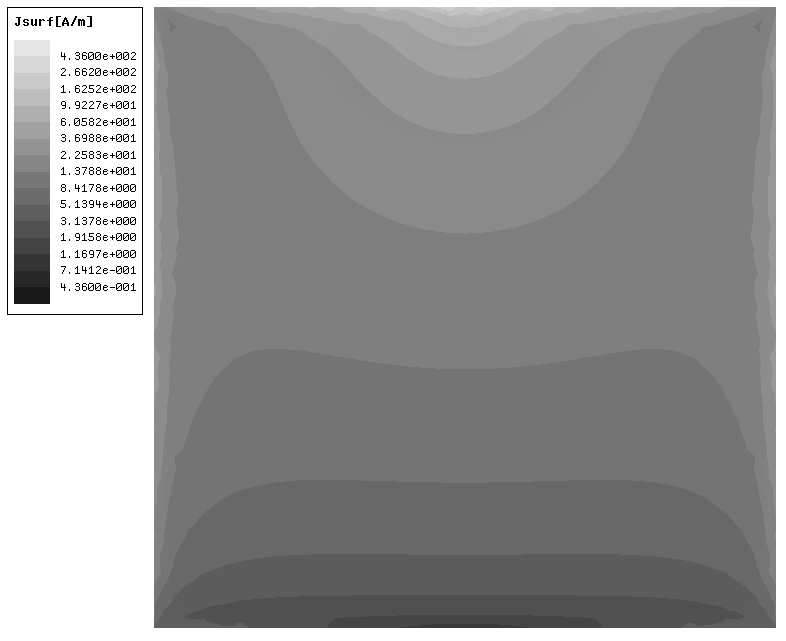}
    \label{fig:pila_magn_jsurf}
  }
  \caption{The surface current at the patch level with a)~dielectric loading with $\epsilon_\lt{r}=4.0$ and b)~magnetic loading with $\mu_\lt{r}=3.5$, both at 550~MHz. The feed is at the center of the top edege of the patch. The surface current at the patch slightly larger with the magnetic loading.}
  \label{fig:pila_jsurf}
\end{figure}

\begin{figure}[!t]
	\centering
	\includegraphics[width=85mm]{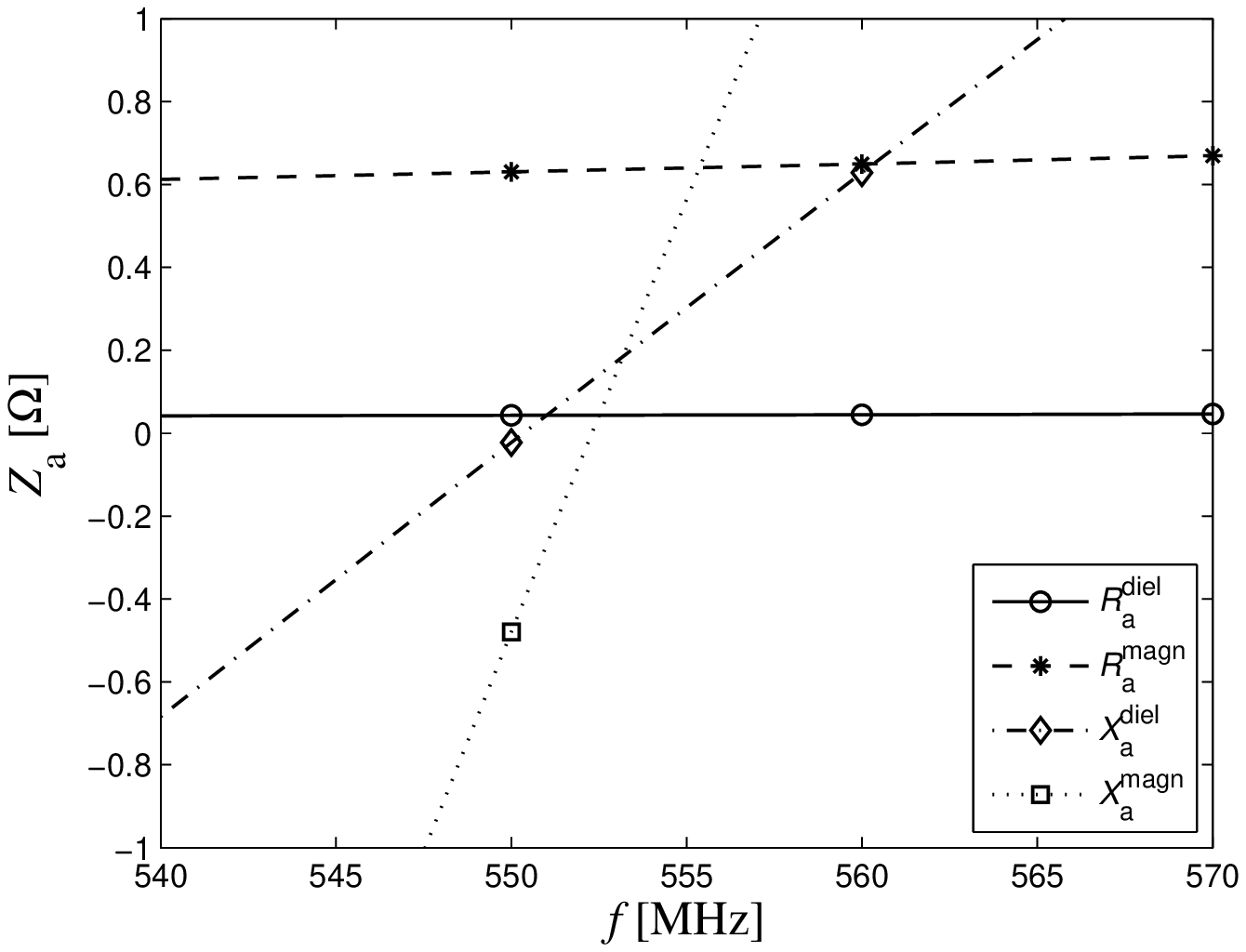}
	\caption{The real and imaginary part of the antenna impedance $Z_\lt{a}$. $R_\lt{a}^\lt{diel}$ and $R_\lt{a}^\lt{magn}$ are the antenna resistances for dielectric and magnetic loading and $X_\lt{a}^\lt{diel}$ and $X_\lt{a}^\lt{magn}$ the reactances, respectively.}
	\label{fig:za_example}
\end{figure}

We can conclude that the magnetic loading can be really useful also for microstrip antennas with a series resonance, at least when the losses can be neglected. Also, it is seen from this example that the resonance type of the antenna impedance as such is not enough to characterize resonant antennas for either magnetic or dielectric loading.

\subsection{Guidelines}

In the above examples we have considered quite simple antennas in which one of the equivalent surface currents strongly dominates in its contribution to radiation. However, in complex antennas may radiate from both of the equivalent currents. For example, a PIFA with a considerable height may radiate also through the shorting element, usually a pin or a narrow strip, due to the high current in this element. In this case, the radiation from the vertical electric current may become comparable with the radiation via the electric field at the open end of the patch. By following our suggestions, the most beneficial loading material would actually be a combination of both dielectric and magnetic materials. In reality, one should of course use the filling materials at the position where the effect is the best, as it is presented in \cite{Luukkonen2007} for partial dielectric loading of microstrip antennas with the $\lambda/2$ resonance. If an antenna radiates from both equivalent currents, partial material loading using both types of substrates may prove most beneficial.

The frequency dispersion of natural magnetic materials is also an important limitation. The benefit of the magnetic loading has to be significant because the dispersion increases the stored energy and also the material losses reduce the radiation efficiency, which has to be taken into account in lossy antennas when calculating the radiation quality factor. Also, the conductor losses are significant when the antenna itself is small enough compared to the free-space wavelength and may be different with the same antenna utilizing dielectric or magnetic materials due to the different magnetic-field distributions inside the antenna.

Dielectric and magnetic substrates as such are hardly equal in their potential for antenna miniaturization. Even when a magnetic material might provide lower radiation quality factor, the radiation efficiency due to the material losses may be lower than with the compared dielectric material filling, and the comparison must be made between the obtained impedance bandwidth and efficiency. Also, the price of magnetic materials due to the manufacturing plays maybe the most restricting role in industrial applications.

\section{Conclusion}

We have introduced a simple and intuitive rule based on the analysis of the radiating fields instead of conventional $RLC$ equivalent circuits or transmission-line theory for determining the beneficial filling material type for small resonant antennas. The method can be applied to antennas radiating through electric and magnetic equivalent currents. The counterparts for these are normal dipole and patch antennas, and the ideal models for these can be assumed to work at qualitative level when applying dielectric and magnetic materials, respectively. An example of quarter-wavelength patch antenna was presented, where the miniaturization can be explained by using our method, whereas the conventional circuit approach would lead to a wrong conclusion about the optimal material loading.

\section{Acknowledgments}

This research was funded in part by Intel Corporation and Nokia Corporations and supported by the Academy of Finland and Nokia through the Center-of-Excellence program.


\bibliographystyle{./IEEEtranBST/IEEEtran}
\bibliography{./IEEEtranBST/IEEEabrv,./karilainen}

\end{document}